\documentclass[a4paper,12pt]{article}

 \usepackage{amsfonts}
\usepackage{verbatim}

\begin{document}

\begin{flushright}
arXiv: 0709.2277 [hep-th]\\ CAS-BHU/Preprint
\end{flushright}

\vskip 0.9cm

\begin{center}

{\small \bf NILPOTENT SYMMETRY
INVARIANCE IN THE NON-ABELIAN 1-FORM GAUGE THEORY: SUPERFIELD FORMALISM}\\

\vskip 2cm

{\bf  R. P. Malik\footnote{Presently
also associated with the DST-Centre for Interdisciplinary 
Mathematical Sciences, Faculty of Science, Banaras Hindu University, Varanasi - 221 005,
(U. P.), India.} and B. P. Mandal} \\ {\it Physics Department, Centre of
Advanced Studies,}\\ {\it Banaras Hindu
University, Varanasi-221 005, (U. P.), India}\\ {\small \bf E-mails:
malik@bhu.ac.in; bhabani@bhu.ac.in}\\

\end{center}

\vskip 2.5cm

\noindent {\bf Abstract:} We demonstrate that the nilpotent
Becchi-Rouet-Stora-Tyutin (BRST) and anti-BRST symmetry invariance
of the Lagrangian density of a four (3 + 1)-dimensional (4D)
non-Abelian 1-form gauge theory with Dirac fields can be captured
within the framework of the superfield approach to BRST formalism.
The above 4D theory, where there is an explicit coupling between
the non-Abelian 1-form gauge field and the Dirac fields, is
considered on a (4, 2)-dimensional supermanifold, parameterized by
the bosonic 4D space-time variables and a pair of Grassmannian
variables. We show that the Grassmannian independence of the super
Lagrangian density, expressed in terms of the (4, 2)-dimensional
superfields, is a clear signature of the presence of the
(anti-)BRST invariance in the original 4D theory.
\\



\newpage

\section{Introduction}

The geometrical interpretations of the nilpotency and
anticommutativity properties, associated with the (anti-)BRST
symmetry transformations of the gauge and (anti-)ghost fields of a
given 1-form gauge theory, have been provided within the framework
of the usual superfield approach to BRST formalism [1-9]. This
approach, however, has not been able to shed any light on the nilpotent
(anti-)BRST symmetry transformations of the matter fields of an
interacting (non-)Abelian 1-form gauge theory.

In a set of research papers [10-20], the above superfield
formulation [1-9], has been consistently extended so as to derive
the nilpotent (anti-)BRST symmetry transformations for the matter
(i.e. Dirac, complex scalar, etc.) fields of a given (non-)Abelian
1-form gauge theory. In the above attempts [1-20], however, the
(anti-)BRST invariance of the Lagrangian densities has {\it not yet} been
captured within the framework of the superfield formulation.

The central theme of a couple of very recent papers [21,22]
concerns itself with the (anti-)BRST invariance of the 4D
(non-)Abelian 1-form gauge theories within the framework of the
superfield approach to BRST formalism. To be precise, it has been
shown that the Grassmannian independence of the super Lagrangian
density, expressed in terms of the superfields defined on the (4,
2)-dimensional supermanifold, is a clear-cut signature of the
presence of the (anti-)BRST invariance in the above 1-form gauge
theories.

The field theoretical models that have been considered in [21,22]
are (i) the 4D (non-)Abelian gauge theory without any interaction
with matter fields [21], and (ii) the interacting Abelian U(1) 1-form
gauge theory with Dirac fields [22]. The purpose of our present
investigation is to generalize our results of [21,22] to the case
of the interacting 4D non-Abelian gauge theory (with Dirac
fields). Furthermore, we demonstrate that, like our earlier
observations in [21,22], the presence of the (anti-)BRST
invariance in the 4D ordinary non-Abelian 1-form gauge theory is
encoded in the proof that the Grassmannian derivatives, acting on
the super Lagrangian density, produce zero result.

The motivating factors that have propelled us to pursue our
present investigation are as follows. First and foremost, it is
important to check the validity of  our results (that were
obtained for the (non-)interacting Abelian 1-form gauge theory) to
our present model of the interacting non-Abelian theory. Second,
it is interesting to explore the geometrical interpretation of the
(anti-)BRST invariance for our present {\it interacting}
non-Abelian gauge theory which were found to be quite cute and
simple for the interacting U(1) gauge theory (with Dirac fields).
Finally, our earlier works [21,22] and our present endeavour are
our modest steps towards our main goal of applying the superfield
formulation to the 2-form (and still higher form) gauge theories.

The material of our present investigation is organized as follows.
In Sec. 2, we recapitulate the bare essentials of the (anti-)BRST
invariance of our present interacting non-Abelian theory in the
Lagrangian formulation. Section 3 is devoted to a brief discussion
of the horizontality condition and its consequences within the
framework of the superfield formulation. Our Sec. 4 deals with a
gauge invariant restriction (GIR) on the matter superfields of the
(4, 2)-dimensional supermanifold and its outcome for the
(anti-)BRST invariance in our present 4D theory. In Sec. 5, we
provide a concise discussion of a {\it single} GIR on the matter
superfields that leads to the consequences of our Secs. 3 and 4.
Finally, we make some concluding remarks in Sec. 6.

\section{Off-shell nilpotent (anti-)BRST symmetry
 invariance: Lagrangian formulation}

Let us begin with the following Lagrangian densities for the 4D
interacting non-Abelian 1-form gauge theory with Dirac fields in
the Feynman gauge [23]
\begin{eqnarray} {\cal L}^{(1)}_b &=& - \frac{1}{4}
F^{\mu\nu} \cdot F_{\mu\nu} + \bar \psi \;(i \gamma^\mu D_\mu -
m)\; \psi + B \cdot (\partial_\mu A^\mu) \nonumber\\ &+&
\frac{1}{2} \;(B \cdot B + \bar B \cdot \bar B) - i
\partial_\mu \bar C \cdot
D^\mu C,
\end{eqnarray}
\begin{eqnarray} {\cal L}^{(2)}_{\bar b} &=& - \frac{1}{4}
F^{\mu\nu} \cdot F_{\mu\nu} + \bar \psi \;(i \gamma^\mu D_\mu -
m)\; \psi - \bar B  \cdot (\partial_\mu A^\mu) \nonumber\\ &+&
\frac{1}{2}\; (B \cdot B + \bar B \cdot \bar B) - i D_\mu \bar C
\cdot
\partial^\mu C,
\end{eqnarray}
where $D_\mu C = \partial_\mu C + i A_\mu \times C$ and $D_\mu
\psi = \partial_\mu \psi + i (A_\mu \cdot T) \psi$ are the
covariant derivatives on the fermionic ghost field $C$ and matter
(Dirac) field $\psi$, respectively. These covariant derivatives
satisfy $ [ D_\mu, D_\nu ] \psi  = i F_{\mu\nu} \psi, [D_\mu,
D_\nu ] C = i F_{\mu\nu} \times C$ which define \footnote{We adopt
here the conventions and notations such that the 4D flat metric
$\eta_{\mu\nu}$ has the signature $(+1, -1, -1, -1)$ and dot and
cross products between two vectors $P^a$ and $Q^a$ in the Lie
algebraic group space are $P \cdot Q = P^a Q^a$ and $(P \times
Q)^a = f^{abc} P^b Q^c$ where $f^{abc}$ are the structure
constants in the $SU(N)$ Lie algebra $[T^a, T^b ] = f^{abc} T^c$
obeyed by the generators $T^a$'s. The latter are present in the
definition of the non-Abelian 1-form $A^{(1)} = dx^\mu A_\mu \cdot
T$. In the above, we have: $\mu, \nu....= 0, 1, 2, 3$ and $a, b, c
= 1, 2, 3...N$.} the curvature tensor $ F_{\mu\nu} =
\partial_\mu A_\nu -
\partial_\nu A_\mu + i A_\mu \times A_\nu$.
Here $B$ and $\bar B$ are the auxiliary fields that satisfy the
Curci-Ferrari condition $B + \bar B = - (C \times \bar C)$ [24] so
as to make the following off-shell nilpotent ($s_{(a)b}^2 =0$)
(anti-)BRST symmetry transformations $s_{(a)b}$ (see, e.g [23])
\begin{eqnarray}
&& s_b A_\mu = D_\mu C, \quad s_b  C =  - \frac{i}{2} (C \times
C), \quad s_b \bar C = i B, \quad s_b B = 0, \nonumber\\ && s_b
\psi = - i (C \cdot T) \psi, \quad s_b \bar \psi = - i  \bar \psi
(C \cdot T), \quad s_b \bar B = i (\bar B \times C),
\end{eqnarray}
\begin{eqnarray}
&& s_{ab} A_\mu = D_\mu \bar C, \quad s_{ab} \bar C = -
\frac{i}{2} (\bar C \times \bar C), \quad s_{ab} C =  i \bar B,
\quad s_{ab} \bar B = 0, \nonumber\\ && s_{ab} \psi = - i  (\bar C
\cdot T)  \psi, \quad s_{ab} \bar \psi = - i  \bar \psi (\bar C
\cdot T), \quad s_{ab} B = i (B \times \bar C),
\end{eqnarray}
anticommutative ($s_b s_{ab} + s_{ab} s_b = 0$) in nature. In the
above, the fields $\bar C^a(C^a)$ are the anticommuting
(anti-)ghost fields that are required for the proof of unitarity
in the theory [25] and $\gamma^\mu$ are the usual $4 \times 4$
Dirac matrices.

The above nilpotent transformations (3) and (4) are the symmetry
transformations because the Lagrangian densities change to total
derivatives under them. The key reasons behind the (anti-)BRST
invariance are (i) the symmetry invariance of the kinetic energy
term (i.e. $s_{(a)b} [F^{\mu\nu} \cdot F_{\mu\nu}] = 0$), (ii) the
invariance of the terms that contain Dirac fields (i.e. $s_{(a)b}
[\bar \psi (i \gamma^\mu D_\mu - m) \psi] = 0$), and (iii) the
invariance of the gauge-fixing and Faddeev-Popov ghost term. In
fact, the last statement can be mathematically expressed as
\begin{eqnarray}
&& s_b s_{ab} \Bigl [ \frac{i}{2} A_\mu \cdot A^\mu + C \cdot \bar
C \Bigr ] =  B \cdot (\partial_\mu A^\mu) + \frac{1}{2} (B \cdot B
+ \bar B \cdot \bar B) - i \partial_\mu \bar C \cdot D^\mu C
\nonumber\\ &&\equiv - \bar B  \cdot (\partial_\mu A^\mu) +
\frac{1}{2}\; (B \cdot B + \bar B \cdot \bar B) - i D_\mu \bar C
\cdot \partial^\mu C.
\end{eqnarray}
The above expression clearly implies (due to the nilpotency and
anticommutativity of $s_{(a)b}$) that the gauge-fixing and
Faddeev-Popov ghost terms {\it together} remain invariant under
the (anti-)BRST symmetry transformations.

\section{Horizontality condition: outcomes}

To tap the potential and power of the celebrated horizontality
condition (HC), first of all, we generalize the ordinary exterior
derivative $d = dx^\mu
\partial_\mu$ and the ordinary 1-form connection
($A^{(1)} = dx^\mu A_\mu)$ of the
4D theory to their counterparts on the (4, 2)-dimensional
supermanifold as follows
\begin{eqnarray}
d \to \tilde d = dx^\mu \;\partial_\mu + d \theta
\;\partial_\theta + d \bar\theta \;\partial_{\bar\theta},
\nonumber\\ A^{(1)} \to \tilde A^{(1)} = dx^\mu \;({\cal B}_\mu
\cdot T) + d \theta \;(\bar {\cal F} \cdot T) + d \bar\theta
\;({\cal F} \cdot T),
\end{eqnarray}
where (${\cal B}_\mu, {\cal F}, \bar {\cal F}$) are the
superfields defined on the above supermanifold. These are the
generalization of the basic fields $(A_\mu, C, \bar C)$ as can be
seen from the following expansion along the Grassmannian
directions [4,5,16]
\begin{eqnarray}
{\cal B}_\mu  (x, \theta, \bar\theta) &=& (A_\mu \cdot T) (x) +
\theta (\bar R_\mu \cdot T) (x) + \bar\theta (R_\mu \cdot T) (x) +
i \theta \bar\theta (S_\mu \cdot T) (x), \nonumber\\ {\cal F} (x,
\theta, \bar\theta) &=&  (C \cdot T)  (x) + i \theta (\bar B \cdot
T) (x) + i \bar\theta  ({\cal B} \cdot T) (x) + i \theta
\bar\theta (s \cdot T) (x), \nonumber\\ \bar {\cal F} (x, \theta,
\bar\theta) &=& (\bar C \cdot T) (x) + i \theta (\bar {\cal B}
\cdot T) (x) + i \bar\theta (B \cdot T) (x) + i \theta \bar\theta
(\bar s \cdot T) (x).
\end{eqnarray}
It is elementary to check that, in the limit $(\theta, \bar\theta)
\to 0$, we retrieve basic 4D local fields $A_\mu, C$ and $\bar
C$ of our Lagrangian densities (1) and/or (2).

The HC is the requirement that the super 2-form $\tilde F^{(2)} =
\tilde d \tilde A^{(1)} + i \tilde A^{(1)} \wedge \tilde A^{(1)}$
is equal to the ordinary 2-form $F^{(2)} = d A^{(1)} + i A^{(1)}
\wedge A^{(1)}$. This equality leads to the determination of the
secondary fields $R_\mu, \bar R_\mu, S_\mu, {\cal B}, \bar {\cal
B}, s , \bar s$ of the expansion (7) in terms of the basic fields
(see, e.g. [16]). The ensuing expansion, with these values
suitably inserted into (7), looks as \footnote{For the sake of
brevity, we use the notations $A_\mu \equiv A_\mu \cdot T, C = C
\cdot T$, etc.}
\begin{eqnarray}
{\cal B}^{(h)}_\mu (x, \theta, \bar\theta) &=& A_\mu  + \theta
\;D_\mu \bar C  + \bar\theta \; D_\mu C  + i\; \theta
\;\bar\theta\; (D_\mu B  + D_\mu C \times \bar C) \nonumber\\
&\equiv& A_\mu (x) + \theta (s_{ab} A_\mu (x)) + \bar \theta (s_b
A_\mu (x)) + \theta \bar\theta\; (s_b s_{ab} A_\mu (x)),
\nonumber\\ {\cal F}^{(h)} (x, \theta, \bar\theta) &=&  C  + i\;
\theta \; \bar B - \frac{i}{2}\; \bar \theta (C \times C) - \theta
\;\bar \theta \;(\bar B \times C) \nonumber\\ &\equiv& C (x) +
\theta\; (s_{ab} C (x)) + \bar\theta\; (s_b C (x)) + \theta
\;\bar\theta \;(s_b s_{ab} C (x)), \nonumber\\ \bar {\cal F}^{(h)}
(x, \theta, \bar\theta) &=& \bar C - \frac{i}{2}\; \theta \;(\bar
C \times \bar C) + i \;\bar\theta\; B  + \theta \;\bar\theta \;(B
\times \bar C) \nonumber\\ &\equiv& \bar C (x) + \theta\; (s_{ab}
\bar C (x)) + \bar\theta\; (s_b \bar C (x)) + \theta \;\bar\theta
\;(s_b s_{ab} \bar C (x)).
\end{eqnarray}
In the above, the superscript $(h)$ stands for the superfields
that are obtained after the application of the HC and $s_{(a)b}$
are the transformations (3) and (4).

It is evident that the (anti-)BRST symmetry transformations can be
expressed as: $s_b \Omega = \mbox{Lim}_{\theta \to 0}
(\partial/\partial \bar\theta) \tilde \Omega^{(h)}, s_{ab} \Omega
= \mbox{Lim}_{\bar\theta \to 0} (\partial/\partial \theta) \tilde
\Omega^{(h)}$. Here the local 4D generic field is $\Omega (x)$ and
its counterpart on the (4, 2)-dimensional supermanifold is the
superfield $\tilde \Omega^{(h)}$ (obtained after the application
of HC). The above mapping provides the geometrical interpretation
of the nilpotent (anti-)BRST transformations $s_{(a)b}$ as the
translational generators $(\partial_\theta,
\partial_{\bar\theta})$ along the Grassmannian directions of the (4,
2)-dimensional supermanifold.

The 2-form super curvature tensor $\tilde F_{\mu\nu}^{(h)} =
\partial_\mu {\cal B}_\nu^{(h)} -
 \partial_\nu {\cal B}_\mu^{(h)}  + i {\cal B}_\mu^{(h)} \wedge
 {\cal B}_\nu^{(h)}$ can be explicitly expanded along the Grassmannian
 directions as [4]
 \begin{equation}
\tilde F_{\mu\nu}^{(h)} = F_{\mu\nu} + i \theta (F_{\mu\nu} \times
\bar C)+ i \bar\theta (F_{\mu\nu} \times C) - \theta \bar\theta
(F_{\mu\nu} \times B + F_{\mu\nu} \times C \times \bar C).
 \end{equation}
The above equation immediately implies that the kinetic energy
term of the Lagrangian densities (1) and/or (2) remains unaffected
due to presence of the Grassmannian variables, namely;
(see, e.g. [4] for details)
\begin{equation}
- \frac{1}{4} \; \tilde F^{\mu\nu(h)} \cdot \tilde
F_{\mu\nu}^{(h)} = - \frac{1}{4} F^{\mu\nu} \cdot F_{\mu\nu}.
\end{equation}
In the above proof, it is the structure constants $f^{abc}$ (which
are chosen to be totally antisymmetric [23] for the SU(N) group)
play an important role. Physically, the above equality shows that
the l.h.s. is, ultimately, independent ($\partial_\theta [\tilde
F^{\mu\nu(h)} \cdot \tilde F_{\mu\nu}^{(h)}] = 0,
\partial_{\bar\theta} [\tilde F^{\mu\nu(h)} \cdot
\tilde F_{\mu\nu}^{(h)}] = 0$) of the Grassmannian variables
$\theta$ and $\bar\theta$. This observation, in turn, implies the
(anti-)BRST invariance of the 4D kinetic energy term in the
framework of superfield approach to BRST formalism because of the
mappings: $s_b \Leftrightarrow \mbox{Lim}_{\theta \to 0}
\partial_{\bar\theta}, s_{ab} \Leftrightarrow
\mbox{Lim}_{\bar\theta \to 0} \partial_\theta$.

In an exactly similar fashion, it can be checked that the
gauge-fixing and Faddeev-Popov ghost terms of the theory (cf. (5))
can be expressed in terms of the Grassmannian derivatives
$(\partial_\theta, \partial_{\bar\theta})$ and the superfields
(obtained after the application of HC) as
\begin{equation}
s_b s_{ab} \Bigl [ \frac{i}{2} A_\mu \cdot A^\mu + C \cdot \bar C
\Bigr ] = \frac{\partial}{\partial\bar\theta}\; \frac{\partial}
{\partial\theta} \Bigl [ \frac{i}{2} {\cal B}^{(h)}_\mu \cdot
{\cal B}^{\mu(h)} + {\cal F}^{(h)} \cdot \bar {\cal F}^{(h)} \Bigr
],
\end{equation}
where the superfields, with superscript $(h)$, are listed in (8).
It is now elementary to the check that the following super
Lagrangian density ($\tilde {\cal L}_M$), containing kinetic
energy, gauge-fixing and Faddeev-Popov ghost terms, namely;
\begin{equation}
\tilde {\cal L}_M = - \frac{1}{4} \; \tilde F^{\mu\nu(h)} \cdot
\tilde F_{\mu\nu}^{(h)} + \frac{\partial}{\partial\bar\theta}\;
\frac{\partial} {\partial\theta} \Bigl [ \frac{i}{2} {\cal
B}^{(h)}_\mu \cdot {\cal B}^{\mu(h)} + {\cal F}^{(h)} \cdot \bar
{\cal F}^{(h)} \Bigr ],
\end{equation}
is the counterpart of its 4D analogue that is represented by the
following Lagrangian density (i.e. ${\cal L}_M$)
\begin{eqnarray}
&& {\cal L}_M = - \frac{1}{4} F^{\mu\nu} \cdot F_{\mu\nu} - \bar B
\cdot (\partial_\mu A^\mu) + \frac{1}{2}\; (B \cdot B + \bar B
\cdot \bar B) - i D_\mu \bar C \cdot
\partial^\mu C\nonumber\\ &&\equiv - \frac{1}{4} F^{\mu\nu} \cdot F_{\mu\nu}
+  B \cdot (\partial_\mu A^\mu) + \frac{1}{2}\; (B \cdot B + \bar
B \cdot \bar B) - i \partial_\mu \bar C \cdot D^\mu C.
\end{eqnarray}
The above Lagrangian density is a part of the Lagrangian densities
(1) and (2). One of the decisive consequences of the HC is that
the super Lagrangian density $\tilde {\cal L}_M$ is independent of
the Grassmannian variables because $\mbox{Lim}_{\theta \to 0}
\partial_{\bar\theta} \tilde {\cal L}_M = 0$ and
$\mbox{Lim}_{\bar\theta \to 0}
\partial_{\theta} \tilde {\cal L}_M = 0$. This key statement is
equivalent to the (anti-)BRST invariance of the kinetic energy,
gauge-fixing and Faddeev-Popov ghost terms of the 4D Lagrangian
density of our present theory.

Mathematically, the above correspondence can be succinctly
expressed as
\begin{equation}
\mbox{Lim}_{\theta \to 0} \frac{\partial}{\partial\bar\theta}
\tilde {\cal L}_M = 0 \Leftrightarrow s_b {\cal L}_M = 0, \quad
\mbox{Lim}_{\bar\theta \to 0} \frac{\partial}{\partial\theta}
\tilde {\cal L}_M = 0 \Leftrightarrow  s_{ab} {\cal L}_M = 0.
\end{equation}
This mapping captures, in a very simple manner, the (anti-)BRST
invariance of the kinetic energy, gauge-fixing and Faddeev-Popov
ghost terms of the Lagrangian density within the framework of the
superfield formalism. In other words, if the action of the
Grassmannian derivatives on the super Lagrangian density happens
to be zero, the corresponding 4D Lagrangian density would respect
the (anti-)BRST invariance. In the above proof (cf. (14)), the
nilpotency (i.e. $\partial_\theta^2 = 0,
\partial_{\bar\theta}^2 = 0$) and anticommutativity (i.e.
$\partial_\theta \partial_{\bar\theta} + \partial_{\bar\theta}
\partial_\theta = 0$) of the translational generators along Grassmannian
directions play key roles.

\section{Gauge invariant restriction: consequences}

To obtain the (anti-)BRST symmetry transformations for the matter
fields of the theory, we exploit the following gauge invariant
restriction (GIR) on the matter superfields [16]
\begin{equation}
\bar \Psi (x,\theta,\bar\theta)\; [ \tilde d + i \tilde A^{(1)(h)}
]\; \Psi (x,\theta,\bar\theta) = \bar \psi (x)\; [ d + i A^{(1)} ]
\;\psi (x),
\end{equation}
where $\tilde A^{(1)(h)} = dx^\mu {\cal B}^{(h)}_\mu + d \theta
\bar {\cal F}^{(h)} + d \bar\theta {\cal F}^{(h)}$ is the super
1-form connection expressed in terms of the superfields listed in
(8). It is interesting to note that, in the above unique
relationship, the HC and matter (super) fields are intertwined in
a gauge (i.e. BRST) invariant manner.

The matter superfields ($\Psi, \bar \Psi$) are the generalizations
of the 4D Dirac fields ($\psi, \bar\psi)$ of the Lagrangian
densities (1) and/or (2) as can be seen from the following
expansion
\begin{eqnarray}
\Psi (x, \theta, \bar\theta) &=& \psi (x) + i  \theta (b_1 \cdot
T) + i \bar\theta (\bar b_1 \cdot T) + i \theta \bar\theta (f
\cdot T), \nonumber\\ \bar \Psi (x, \theta, \bar\theta) &=& \bar
\psi (x) + i  \theta (b_2 \cdot T) + i \bar\theta (\bar b_2 \cdot
T) + i \theta \bar\theta (\bar f \cdot T),
\end{eqnarray}
where the secondary fields $b_1, \bar b_1, b_2, \bar b_2, f , \bar
f$ are determined in terms of the basic fields from the GIR
(15). These expressions are as follows [16,19]
\begin{eqnarray}
&& b_1 = - (C \cdot T) \psi, \; \bar b_1 = - (\bar C \cdot T)
\psi, \; b_2 = - \bar \psi (C \cdot T), \; \bar b_2 = - \bar\psi
(\bar C \cdot T), \nonumber\\ && f = - i \;[B + \frac{1}{2} (C
\times \bar C) ]\; \psi, \quad \bar f = + i\; \bar \psi\; [ B +
\frac{1}{2} (C \times \bar C) ].
\end{eqnarray}
Insertions of the above values into the expansion (16) leads to
[16-19]
\begin{eqnarray}
\Psi^{(G)} (x, \theta, \bar\theta) &=& \psi(x) + \theta (-i \bar C
\cdot T \psi) + \bar\theta (- i C \cdot T \psi) + \theta
\bar\theta (B + \frac{1}{2} C \times \bar C) \psi, \nonumber\\
&\equiv& \psi (x) + \theta (s_{ab} \psi (x)) +  \bar\theta (s_b
\psi (x)) + \theta \bar\theta (s_b s_{ab} \psi (x)), \nonumber\\
\bar \Psi^{(G)} (x, \theta, \bar\theta) &=& \bar \psi + \theta (-
i \bar \psi \bar C \cdot T) + \bar \theta (-i \bar \psi C \cdot T)
- \theta \bar\theta \bar \psi (B + \frac{1}{2} C \times \bar C),
\nonumber\\ &\equiv& \bar \psi (x) +   \theta (s_{ab} \bar \psi
(x)) + \bar\theta (s_b \bar \psi (x)) +  \theta \bar\theta (s_b
s_{ab} \bar\psi (x)),
\end{eqnarray}
where the nilpotent transformations $s_{(a)b}$ are
listed in (3) and (4) and the superscript $(G)$ on the above matter
superfields denotes that these superfields have been obtained after the 
application of GIR.

As a consequence of the above expansion (18) (that has been
obtained after the application of the GIR in (15)), it is clear
that the following equality (that would be useful for our
discussions) is true, namely;
\begin{equation}
\bar \Psi^{(G)} (x, \theta, \bar\theta) \;  \Psi^{(G)} (x, \theta,
\bar\theta) = \bar \psi (x) \psi (x).
\end{equation}
Furthermore, it can be checked that the following key equality is
also valid on the matter superfields (after the application of GIR
and HC):
\begin{equation}
\bar \Psi^{(G)} [ i \gamma^M D^{(h)}_M - m ] \Psi^{(G)} = \bar
\psi (x) \;(i \gamma^\mu D_\mu - m)\; \psi (x) \equiv {\cal L}_d,
\end{equation}
where ${\cal L}_d$ is the 4D Lagrangian density that contains
Dirac fields and $\gamma^M$ are the generalizations of the $4
\times 4$ Dirac matrices onto (4, 2)-dimensional supermanifold.
With the specific choice of $\gamma^M = (\gamma^\mu, C_\theta,
C_{\bar\theta})$, we obtain
\begin{equation}
\gamma^M D_M^{(h)} = \gamma^\mu (\partial_\mu + i {\cal
B}^{(h)}_\mu) + C_\theta\; (\partial_\theta + i \bar {\cal
F}^{(h)}) + C_{\bar\theta} \; (\partial_{\bar\theta} + i {\cal
F}^{(h)}),
\end{equation}
where $C_\theta$ and $C_{\bar\theta}$ are some anticommuting
($C_\theta^2 = 0, C_{\bar\theta}^2 = 0, C_\theta C_{\bar\theta} +
C_{\bar\theta} C_\theta = 0$) constants which go to zero (i.e.
$C_\theta \to 0, C_{\bar\theta} \to 0$) in the limiting case of
$(\theta, \bar\theta) \to 0$. These requirements on $C_\theta$ and
$C_{\bar\theta}$ are essential so as to  maintain the bosonic
nature of the r.h.s. and to prove that:
\begin{equation}
\mbox{Lim}_{(\theta, \bar\theta) \to 0} \gamma^M D_M = \gamma^\mu
D_\mu \equiv \gamma^\mu (\partial_\mu + i A_\mu \cdot T).
\end{equation}
The above equation implies that, ultimately, we obtain the
ordinary 4D Dirac Lagrangian density when the Grassmannian
variables are set equal to zero.

The exact mathematical form of the constants $C_\theta$ and
$C_{\bar\theta}$ is {\it not} important for our present
discussions because, irrespective of their form, the following
equations (with the superfields ${\cal F}^{(h)}, \bar {\cal
F}^{(h)}, \Psi^{(G)})$ are always satisfied
\begin{equation}
(\partial_\theta + i \; \bar {\cal F}^{(h)})\; \Psi^{(G)} = 0,
\qquad (\partial_{\bar\theta} + i \;  {\cal F}^{(h)})\; \Psi^{(G)}
= 0.
\end{equation}
As a consequence, the exact mathematical form of the anticommuting
constants $C_\theta$ and $C_{\bar\theta}$ does not affect the key
results that emerge from the equation (20) which happens to be a
GIR on the matter superfields of the theory. With inputs from
(23), it is clear that the condition (20) reduces to
\begin{equation}
\bar \Psi^{(G)}\; [ i \gamma^\mu (\partial_\mu + i {\cal
B}_\mu^{(h)}) - m ]\; \Psi^{(G)} = \bar \psi (x) \;(i \gamma^\mu
D_\mu - m)\; \psi (x).
\end{equation}
The above equation is readily satisfied if we insert the
superfield expansions (8) and (18) that have been obtained after
the application of HC and GIR.

It is clear from the equation (24) that the super Lagrangian
density ($\tilde {\cal L}_d$) with gauge and matter superfields
and the ordinary Lagrangian density (${\cal L}_d$) with gauge and
Dirac ordinary fields, namely;
\begin{equation}
\tilde {\cal L}_d = \bar \Psi^{(G)}\; [ i \gamma^\mu (\partial_\mu
+ i {\cal B}_\mu^{(h)}) - m ]\; \Psi^{(G)}, \quad {\cal L}_d =
\bar \psi (x) \;(i \gamma^\mu D_\mu - m)\; \psi (x),
\end{equation}
are equal in the sense that the (4, 2)-dimensional super
Lagrangian density $\tilde {\cal L}_d$ is effectively independent
of the Grassmannian variables $\theta$ and $\bar\theta$. Thus, the
(anti-)BRST invariance can be expressed by the following mappings
\begin{equation}
\mbox{Lim}_{\theta \to 0} \frac{\partial}{\partial\bar\theta}
\tilde {\cal L}_d = 0 \Leftrightarrow s_b {\cal L}_d = 0, \quad
\mbox{Lim}_{\bar\theta \to 0} \frac{\partial}{\partial\theta}
\tilde {\cal L}_d = 0 \Leftrightarrow  s_{ab} {\cal L}_d = 0.
\end{equation}
Here $s_{(a)b}$ are the transformations that are given in (3) and
(4).

 It is worthwhile to recall that the ordinary Lagrangian
density ${\cal L}_d$ remains invariant (i.e. $s_{(a)b} [ \bar \psi
(i \gamma^\mu \partial_\mu - m) \psi = 0$) under the symmetry
transformations $s_{(a)b}$. This is what is reflected in the
Grassmannian independence of the super Lagrangian density (cf.
(26)). In other words, the GIR in (15) leads to the derivation of
a condition (20) which, in turn, implies that the Grassmannian
derivatives acting on the super Lagrangian density $\tilde {\cal
L}_d$ produce zero result.

\section{Single gauge invariant restriction: impacts}

To obtain all the results of Secs. 3 and 4, we begin with the
following GIR on the matter superfields (see, e.g. [19])
\begin{equation}
\bar \Psi (x,\theta,\bar\theta)\; \tilde D \;\tilde D\; \Psi
(x,\theta,\bar\theta) = \bar \psi(x)\; D\; D \;\psi (x)
\end{equation}
where the (super) covariant derivatives ($\tilde D) D$ and their
very intimate connection with the (super) 2-forms $(\tilde
F^{(2)}) F^{(2)}$ are intertwined together in a beautiful manner.
In the above, the 1-form covariant derivatives are defined as
\begin{equation}
\tilde D = \tilde d + i \;\tilde A^{(1)}, \;\;\qquad \;\;D = d +
i\; A^{(1)},
\end{equation}
where all the symbols have been explained in our previous
sections.

It should be noted that the above restriction is also gauge
invariant because the r.h.s. can be explicitly expressed as
\begin{equation}
\bar \psi (x)\; D\; D \;\psi (x) = i \bar \psi (x) F^{(2)} \psi
(x),
\end{equation}
where $F^{(2)} = \frac{1}{2} (dx^\mu \wedge dx^\nu) [ \partial_\mu
A_\nu - \partial_\nu A_\mu + i A_\mu \times A_\nu ]$. It is clear
that, under the SU(N) gauge transformations $\psi \to U \psi, \bar
\psi \to \bar\psi U^{-1}, F^{(2)} \to U F^{(2)} U^{-1}$, the above
expression remains invariant. Here $U \in SU(N)$ is the Lie
algebraic (group valued) unitary transformations on the Dirac
fields as well as SU(N) gauge field. The latter, in turn, implies
the transformation for $F^{(2)}$.

The points to be emphasized, at this stage, are as follows. First,
we obtain all the results that have been obtained (separately and
independently) due to the applications of HC (cf. Sec. 3) and the
GIR (cf. (15) and (20)) in one stroke from our single GIR in (27).
Second, our unique relation (27) combines the (super) curvature
2-forms $(\tilde F^{(2)})F^{(2)}$, (super) covariant derivatives
$(\tilde D) D$ and matter (super) fields in  a beautiful manner.
Finally, it is gratifying that the super curvature tensor $\tilde
F_{\mu\nu}$, that has Grassmannian dependence under HC (cf. (9)),
is now free of them (see, e.g. [19] for details). As a result, one
need not exploit the total super kinetic energy term to show the
Grassmannian independence of the latter. Thus, our restriction
(27) provides an alternative to (and generalization of) the HC as
well as GIR in (15).

Ultimately, it can be seen that the total super Lagrangian density
$\tilde {\cal L}_T$, defined in terms of the (4, 2)-dimensional
superfields, can be expressed as
\begin{equation}
\tilde {\cal L}_T = \tilde {\cal L}_M + \tilde {\cal L}_d,
\end{equation}
where the symbols have been explained earlier in Secs. 3 and 4.
The (anti-)BRST invariance of the 4D theory can be captured in the
language of the total super Lagrangian density ($\tilde {\cal
L}_T$) and the Grassmannian derivatives as:
\begin{equation}
\mbox{Lim}_{\theta \to 0} \frac{\partial}{\partial\bar\theta}
\tilde {\cal L}_T = 0 \Leftrightarrow s_b {\cal L}_T = 0, \quad
\mbox{Lim}_{\bar\theta \to 0} \frac{\partial}{\partial\theta}
\tilde {\cal L}_T = 0 \Leftrightarrow  s_{ab} {\cal L}_T = 0.
\end{equation}
Thus, we note that the real impact of the restriction (27) on the
superfields, defined on the (4, 2)-dimensional supermanifold, is
the Grassmannian independence of the total super Lagrangian
density $\tilde {\cal L}_T$.

\section{Conclusions}

One of the highlights of our present investigation is the
simplicity that has been brought into the discussion of the
(anti-)BRST invariance in the context of the 4D non-Abelian 1-form
gauge theory (with Dirac fields). All one has to basically show is
the Gassmannian independence of the (4, 2)-dimensional super
Lagrangian density of the theory expressed in terms of the
superfields that are obtained after the application of the HC and
GIR.

 Geometrically, the
following points are important for the existence of the
(anti-)BRST invariance within the framework of the superfield
formulation. First, if the translation of the super Lagrangian
density along the $\bar\theta$-direction of the (4, 2)-dimensional
supermanifold is zero, there will be BRST invariance in the 4D
theory. Second, if the above statement is valid for the
$\theta$-direction of the supermanifold, there will be anti-BRST
invariance in the theory. Finally, if the above statements are
valid for the both the Grassmannian directions {\it together},
there will be (anti-)BRST invariance together in the theory.

A very interesting feature of the superfield approach to BRST
formalism is as follows. There is an absolute certainty that the
(anti-)BRST symmetry transformations $s_{(a)b}$ would always be
nilpotent and anticommuting as, geometrically, these correspond to
the translational generators $(\partial_\theta,
\partial_{\bar\theta})$ along the Grassmannian directions of the
(4, 2)-dimensional supermanifold. The latter have the natural
property that $\partial_\theta^2 = 0,
\partial_{\bar\theta}^2 = 0$ and $\partial_\theta
\partial_{\bar\theta} + \partial_{\bar\theta} \partial_\theta =
0$. Thus, the above two key properties of the (anti-)BRST
symmetries are always encoded (and in-built) in our present
superfield approach to BRST formalism.

In a very recent publication of one of us [26], an absolutely
anticommuting (anti-)BRST symmetry transformations have been
obtained in the context of the 4D Abelian 2-form gauge theory
where the superfield approach to BRST formalism has played a key
(but somewhat hidden) role. In this attempt, it has also been
shown that the anticommutativity property of the (anti-)BRST
transformations is deeply connected with the concepts of gerbes.

One of us has been involved with a slightly different type of superspace
(also called the BRST superspace) formulation which has also been 
applied to study gauge theories [27-29]. The central feature of this 
approach is that the whole super Lagrangian density has been accommodated
in a single compact (4, 2)-dimensional gauge invariant action from which the
WT identities emerge very naturally. As a consequence, this type of superspace 
formulation is useful in studying the renormalization of gauge theories.

It would be interesting to unify both the above types of superfield 
approaches to BRST formalism and study the 4D
and 6D (non-)Abelian 2-form gauge theories. In particular, the
application of our superspace formulation to the (higher-form) 
tensor gauge field theories is quite attractive. We are intensively involved, at
present, with the above promising problems and we plan to report about these
developments in our future publications.

\end{document}